\documentclass[conference]{IEEEtran}
\title{Malaria Detection and Classificaiton}
\author{ 
\IEEEauthorblockN{Ruskin Raj Manku, Ayush Sharma, Anand Panchbhai}
\IEEEauthorblockA{Indian Institute of Technology, Bhilai\\
Email: \{ruskinr, ayushsharma, anandp\}@iitbhilai.ac.in
}
}
\usepackage{listings}
\usepackage[utf8]{inputenc}
\usepackage[numbers]{natbib}
\usepackage{graphicx}
\usepackage{tikz}
\usepackage{amsmath}
\usetikzlibrary{positioning,arrows.meta}
\begin{document}
\maketitle
\begin{abstract}
Malaria is a disease of global concern according to the  World  Health  Organization.  Billions  of  people  in  the  world are  at  risk  of  Malaria  today.  Microscopy  is  considered  the  gold standard for Malaria diagnosis. Microscopic assessment of blood samples requires the need of trained professionals who at times are not available in rural areas where Malaria is a problem. Full automation  of  Malaria  diagnosis  is  a  challenging  task. In  this  work, we  put  forward  a  framework  for  diagnosis  of malaria. We adopt a two layer approach, where we detect infected cells using a Faster-RCNN in the first layer, crop them out, and feed the cropped cells to a seperate neural network for classification. The  proposed  methodology  was tested on an openly available dataset, this will serve as a baseline for the future methods as currently there is no common dataset on which results are reported for Malaria Diagnosis.
\end{abstract}
\section{Introduction}
Malaria is a life-threatening disease caused by parasites that are transmitted to people through the bites of infected female Anopheles mosquitoes. In 2017, there were an estimated 219 million cases of malaria in 87 countries. The estimated number of malaria deaths stood at 435,000 in 2017 \cite{malaria99}. It is caused by any of the four different species of the
Plasmodium parasite: Plasmodium vivax(Pv), Plasmodium ovale(Po), Plasmodium malariae(Pm) and Plasmodium falciparum(Pf). Malaria is commonly diagnosed by microscopic examination of blood cells using blood films \cite{shitty}. The diagnosis of these blood films depends on the skill of the pathologist and level of parasites present.  As malaria is generally associated with poverty and occurs mostly in low economic countries \cite{income}, most laboratories or diagnostic facilities are not equipped with standard testing facilities.

Automatic parasite counting has several advantages compared with manual counting: (1) it provides a more reliable and standardized interpretation of blood films, (2) it allows more patients to be served by reducing the workload of the malaria field workers, and (3) it can reduce diagnostic costs \cite{advantage}. Thus, the idea of automating malaria diagnosis has its obvious advantages and has attracted several researchers in the last decade. There have been several attempts to identify malaria infected cells using microscopic images acquired from different parasites of different blood smears using various image processing morphological computer vision techniques. There have also been research for automating the task making use of deep neural networks. More details are stated in past work section. Most of these works fail to provide field level results as most these paper give results on giemsa stained smears, but due to expensive nature, it is generally not used in low economic areas where malaria is an alarming issue. In this paper, we propose a model for malaria detection using images of thin blood smears, our model also obtained good results on images of smears stained with field stain which is a inexpensive stain and commonly used by pathologists in low economic areas of India. Section V provides further details on our method.

The key contributions of this paper can be summarized as follows:
\begin{itemize}
\item We adopt a two layer approach and show how it works better than a one layer approach for Malarial cell Detection and Classification.
\item We provide results for an openly avaiable dataset, this will act as a benchmark for future works for fair comparison between methods as currently no such benchmark dataset is available.
\end{itemize}

\section{Past work}

A lot of research has already been conducted for the diagnosis of malaria parasite in thick/thin blood smear. In this section we go through few such ideas and discuss about the methods deployed. Our main focus will be on methods that have been used in case of thin blood smears.

A system to detect parasitic regions from images of giemsa stained thin blood smears using image processing has been proposed in \cite{1}, it mentions about detection of malarial parasites using HSV segmentation and watershed segmentation, it helps to detect possible parasitic regions which  help the pathologists to reduce false positives, still the solution remains time consuming \& the system fails to classify the parasite autonomously.

A number of image processing techniques have been used in \cite{2}. Objects containing parasites,WBC’s and artefacts are extracted first. Then 23 features(like mean,skewness etc) from these images are fed into Neural Net and SVM. The Neural Network achieves an accuracy of 78.53\% and the SVM achieves an accuracy of 98.25\%.

A two-stage detection and classification has been proposed in \cite{3}. They train a faster-RCNN \cite{DBLP:journals/corr/RenHG015} to detect RBC and non-RBC(like trophozoite,ring etc). After the faster-RCNN will give region of interest for non-RBC cells, a separate CNN classifier has been used to classify the non-RBC cells into further classes. This system achieved an accuracy of 98\% . This study lacks emphasis on field effectiveness of their algorithm because of sharp and clear images in the dataset which might not be the case in real life. Although we adopt a similar approach, the key goal of this study is to provide a benchmark on an openly available dataset.

A system to classify malarial parasites in acridine orange (AO) stained blood smears has been proposed in \cite{4}, the proposed method uses connected component analysis to segment pre processed imaged and then uses Ada-Boost to create a classifier , though this provides fast detection, this system uses fluorescent acridine orange staining which rely on fluorescence microscopes, this staining method is generally not available for field level diagnosis.

Malaria parasite detection and species identification on thin blood smears has been proposed in \cite{5}. The dataset used is composed of 363 images. The smears used for the images were obtained during a field study done in Palawan, Philippines. Connected component analysis method has been used to create bounding boxes for cells. Segmented images are then fed into pre-trained Inception V3 , and transfer learning is used to classify the plasmodium species. Running on the test set, the model obtained an accuracy of 92.4\% for parasite detection (with 95.2\% sensitivity and 84.7\% specificity) and 87.9\% in species identification. 

An automated malaria diagnosis system as proposed in \cite{6} uses object detection module to process the images pre-processed with white balancing techniques, CNNs for feature extraction with  gamma-transform color augmentation scheme. This system has implemented field level detection of malaria in a robust manner. This system was built for thick smear slides, thus is only able to detect malaria but not classify the types. 

In \cite{8} they have built a three stage classifier and have categorized the the cells into 13 categories from giemsa stained thin blood smears, including four ring categories, i.e. one for
each species (Pf, Pv, Po, Pm), similarly four transitional and four late stage categories and one for artifacts (or distractors). They have also taken the number of RBCs in consideration to generate elaborated patient-level reports. They have provided metrics at patient level, claiming to have achieved 90\% sample-level specificity on holdout sets. Their model performs well on Pf \& Pv type malaria (achieved accuracy above 90\%) but does not show similar performance for Po \& Pm type malaria.  

A very detailed comparison of techniques used in past has been given in \cite{7}. This study provides comparison of techniques used for both thick and thin smears.

Most of these systems acquired data in a very controlled environment, thus are not very robust in field case scenarios. Also, many of these systems are built for the detection of malaria and fail to classify the specific parasitic class causing the Malaria. 

\section{Problem Statement}
Given a blood slide, we have to detect if the slide contains malaria infected cells. Furthur, we should classify the infected cells: Ring, Trophozoite, Schizont and Gametocyte to provide stage-wise classification results. We evaluate our model on the basis of precision and recall. A good model should be able to give a high precision and recall on slides it has never seen before. This includes slides of a different patient and slides prepared with a different stain than the model is trained on. The model should also work good for slides prepared at field level and not only for high quality slides that are prepared in highly sophisticated labs.

\section{Dataset}

We have used image set BBBC041v1, available from the Broad Bioimage Benchmark Collection\cite{dataset} as our training data. The dataset contains 1364 images (~80,000 cells)of blood smears stained with Giemsa reagent. The data consists of two classes of uninfected cells (RBCs and leukocytes) and four classes of infected cells (gametocytes, rings, trophozoites, and schizonts). Dataset was obtained as collection of 3 sets, with different researchers having prepared each one: from Brazil, from Southeast Asia, and time course.\

\section{Methodology }
We follow a two layer approach in order to do classification at cellular level.
\subsection{Layer 1: Infected cell detection}
Given a blood slide, first we detect any infected cell. We consider the classes Ring, Trophozoite, Schizont and Gametocyte as infected. In order to detect an infected cell we use the state of art object detection model Faster RCNN. The model works as follows. \\The image is provided as an input to a pretrained convolutional network which provides a convolutional feature map. The feature map generated is then fed to a region proposal network(RPN). The RPN gives regional proposals for the objects. The predicted region proposals are then reshaped using a RoI pooling layer which is then used to classify the image within the proposed region and predict the offset values for the bounding boxes.\\
\begin{center}
\includegraphics[scale=0.3]{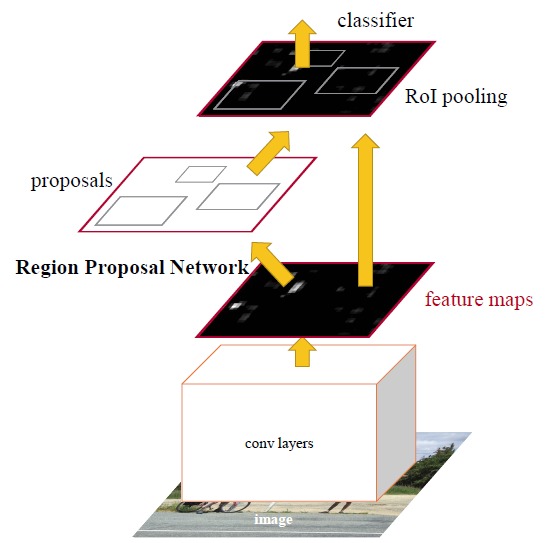}
\end{center}
After we detect infected cells, we cut out those cells and do a cellular level classification in the second layer.
\subsection{Layer 2: Infected cell classification}
We get images of infected cells from layer 1. We run this image through a pretrained ResNet-50 model and get a feature vector for this image. Then we feed this feature vector as input to a 4-layer neural network which performs the task of stage-level classification. \\
We adopt a two layer approach and not a one layer approach that trains a Faster-RCNN for both detection and classication of cell stages. The reason for this is: as mentioned before, Faster-RCNN has a ROI pooling layer. The purpose of this layer is to generate fixed size feature maps for the classification of regions. The layer achieves this by using max-pooling. However, max pooling can lead to loss of image features and subsequently a lower accuracy of classification. We overcome this issue by cropping out the region of interest and reshaping images to a fixed size, allowing us to use images of same size without any loss of image features.\\
In the following section we dive deeper into the models implemented.
\section{Model details}
\subsection{Layer 1: Infected cell detection}
This model is trained to detect infected cells in a given blood sample.
\begin{itemize}
\item We consider the classes Ring, Trophozoite, Schizont and Gametocyte as one single "infected" class.
\item \textbf{1} output class
\item \textbf{15,000} iterations
\item $mAP - \textbf{0.832}$
\item $recall - \textbf{0.69}$\\
\end{itemize}
As evident, this model gave a very high precision while detecting infected cells. We then crop out the bounding boxes for infected cells and pass them through our second layer.

\subsection{Layer 2: Infected cell classification}
We feed the infected cell image we got from layer 1 to a pretrained ResNet-50 model. This model gives a features vector of length 2048. Now we feed this 2048 size vector of all the images to a 4-layer Neural Network. We clip the number of images from each class category to $140$ images in order to have a balanced dataset. Following are the training details of the Neural Network model :-
\begin{itemize}
\item $140$ images each from classes Ring, Trophozoite , Schizont and Gametocyte are used to create the dataset. For a total of $560$ images, $90\%$ images are fed as train and $10\%$ images for testing.
\item Following table describes the classification report of the Neural Network model.
\begin{center}
\begin{tabular}{|c|c|c|c|}
\hline
Class name & precision & recall & f1-score \\
\hline
Gametocyte & 0.80 & 0.80 & 0.80 \\
Ring & 0.91 & 0.77 & 0.83 \\
Schizont & 0.87 & 0.87 & 0.97 \\
Trophozoite & 0.73 & 0.85 & 0.79 \\
\hline
Metric & precision & recall & f1-score \\
\hline
Macro avg & \textbf{0.83} & \textbf{0.82} & \textbf{0.82} \\
Weighted avg & \textbf{0.83} & \textbf{0.82} & \textbf{0.82}\\
\hline
\multicolumn{1}{|c}{} & \multicolumn{1}{c}{} & \multicolumn{1}{c}{Accuracy:} & \multicolumn{1}{c|}{\textbf{0.82}}\\
\hline
\end{tabular}
\end{center}
\end{itemize}
Just as we expected, a two layer approach works better not only theoretically, but also empirically. If we train a Faster-RCNN model to directly predict the classes, i.e if we adopt a one layer approach, the mAP(Mean Average Precision) achieved is \textbf{0.4}, and the average recall is \textbf{0.5}; the cell classification results have improved significantly.

\section{Limitations and future work}
One key limitation of our study is the lack of field testing results; A clear understanding about how good of a precision, recall and accuracy are needed for reliable automatic field testing is key area that needs to be explored in future. There is a lot of variation if one takes into consideration the different types of stains available, thick and thin blood smears, we plan to work on a method that'll be able to handle such variations. Such a model should focus on the morphology of the parasites and should be able to deal with the debris that are often found in images procured from the field. The key to developing such a model could be the use of appropriate pre-processing steps or using domain adaptation techniques like gradient reversal\cite{ganin2015unsupervised}

\section{Conclusion}
We have proposed a two stage classification of infected cells and their classificaiton. We observe how a two layer approach outperforms a single layer approach. Although a two layer approach has been proposed before, such studies have used a very controlled and high quality dataset, not suitable for field testing in rural areas, and those datasets are not available online for fair comparison as well. We report the results on an openly available field level dataset which will act as a baseline for future methods.
\bibliographystyle{unsrt}
\bibliography{paper.bib}

\end{document}